\documentclass[twocolumn,showpacs,preprintnumbers,amsmath,amssymb]{revtex4}

\usepackage{graphicx}
\usepackage{dcolumn}
\usepackage{bm}

\begin{document}

\preprint{APS/123-QED}

\title{On the information-splitting essence of\\ two types of quantum key distribution protocols}

\author{Fei Gao,$^{1,2,}$\footnote{Electronic address: hzpe@sohu.com} Fen-Zhuo Guo,$^{1}$ Qiao-Yan Wen,$^{1}$ and Fu-Chen Zhu$^{3}$}
\affiliation{
$^{1}$School of Science, Beijing University of Posts and Telecommunications, Beijing, 100876, China\\
$^{2}$State Key Laboratory of Integrated Services Network, Xidian
University, Xi'an, 710071, China\\
$^{3}$National Laboratory for Modern Communications, P.O.Box 810,
Chengdu, 610041, China}

\date{\today}

\begin{abstract}
With the help of a simple quantum key distribution (QKD) scheme,
we discuss the relation between BB84-type protocols and
two-step-type ones. It is shown that they have the same essence,
i.e., information-splitting. More specifically, the similarity
between them includes (1) the carrier state is split into two
parts which will be sent one by one; (2) the possible states of
each quantum part are indistinguishable; (3) anyone who obtains
both parts can recover the initial carrier state and then
distinguish it from several possible states. This result is useful
for related scheme designing and security analyzing.
\end{abstract}

\maketitle

The goal of cryptography is to ensure that a secret message is
transmitted between two users, traditionally called Alice and Bob,
in a way that any eavesdropper (such as Eve) cannot read it. Till
now, it is generally accepted that the only proven secure
cryptosystem is the scheme of one-time pad, which utilizes a
previously shared secret key to encrypt the message transmitted in
the public channel. However, it is difficult for all existing
classical cryptosystems to establish a random key with
unconditional security between Alice and Bob. Fortunately, quantum
key distribution (QKD)
\cite{BB84,E91,B92,BW92,GV95,HIGM,KI97,B98,C2000,CHL,LCA,GLSLG,LL02,XLG,GRTZ,ZYCP,GGWZ,D04,W05,G05},
the approach using quantum mechanics principles for the
distribution of secret key, can accomplish this task skillfully.

The first QKD scheme, i.e., BB84 protocol, was proposed in 1984 by
Bennett and Brassard \cite{BB84}. Here we give a brief
introduction of a well-known variation of it, i.e., the delayed
choice BB84 protocol. In this scheme Alice randomly generates
single photons polarized along one of four possible directions,
$0^\circ$, $45^\circ$, $90^\circ$ or $135^\circ$, and sends them
to Bob. Note that these four states form two conjugate bases
$\{0^\circ,90^\circ\}$ and $\{45^\circ,135^\circ\}$. After Bob
received these photons, Alice tells Bob which basis she used for
each one. Afterwards, Bob measures every photon in the same basis
as that of Alice and it follows that he can know what states Alice
sent from the measurement results. As the different states can be
encoded into 0 and 1, Alice and Bob will share a sequence of key
bits. Eve's presence can be detected by publicly comparing a
subsequence of these bits. Obviously, Eve cannot elicit correct
information without introducing errors because she does not know,
when these photons are transmitted in the quantum channel, which
of the two conjugate bases was used to prepare each of them. If
Eve tries to measure one certain photon to acquire information,
she will bring disturbance once she uses a wrong basis (the
announcement of bases from Alice is too late for Eve).
Nevertheless Bob will obtain all of the key bits (noise is not
concerned here) sent by Alice because he can perform his
measurement after Alice's announcement. Several other schemes
utilize similar means to distribute key (for example, see
Refs.\cite{B92,B98}). We customarily call them BB84-type
protocols. The main feature of these protocols is that the
communicators use nonorthogonal states so that Eve cannot
distinguish them completely, on which the security is based.

\begin{figure}
\includegraphics{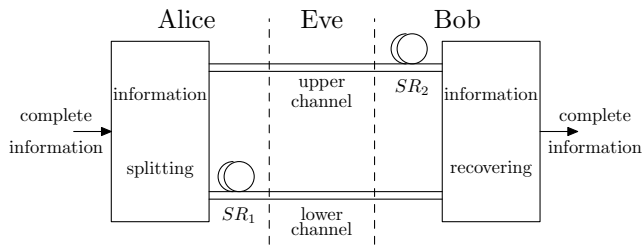}
\caption{\label{fig:one} The QKD model of two-step-type protocols.
In these protocols Alice splits the complete information (carried
by orthogonal states) into two quantum parts and sends them one by
one. That is, they never appear simultaneously in the insecure
channel, which can be achieved by using storage rings $SR_1$. At
the destination, with the help of storage rings $SR_2$, Bob can
recover the complete information after both two parts arrived. In
fact, BB84-type protocols can also be depicted by this model. In
this condition the lower channel is a classical channel instead of
a quantum one. See the following paragraphs for details.}
\end{figure}

Since Goldenberg and Vaidman came up with a QKD protocol based on
orthogonal states in 1995 \cite{GV95}, much attention has been
focused on this type of quantum cryptography
\cite{KI97,GLSLG,LL02,DLL05}. In these schemes the information
carrier is composed of a set of orthogonal states. As we know,
orthogonal states can be easily distinguished with no disturbance
and, therefore, as the carrier, they cannot be transmitted in the
public quantum channel directly. The above proposals give a smart
resolvent, that is, splitting the information unit, which denotes
the information encoded in one complete unit of carrier, into two
parts, e.g., the two parts of a photon wave packet
\cite{GV95,KI97} or two correlated particles
\cite{GLSLG,LL02,DLL05}, and sending them one by one (see
Fig.~\ref{fig:one}). Note that one can never read the total
information transmitted without introducing disturbance by
measuring only one of the two parts. In this condition, Eve cannot
access the two parts simultaneously and it follows that her
eavesdropping is doomed to failure. In this Letter we ignore the
case that Eve intercepts legal wave packets (or particles) and
sends some fake ones to Bob, where Eve will introduce errors
though she can obtain both two parts at the same time. Because the
information unit is always transmitted by two steps, we call this
type of schemes two-step-type protocols.

In this Letter we focus our attention on the relation between
BB84-type protocols and two-step-type ones. In fact, Peres and the
authors of Ref.\cite{GV95} discussed the similar topic long time
ago, but they did not make an agreement about it \cite{P96,GV96}.
As far as the great influence of both types of QKD on quantum
cryptography is concerned, it is of importance to clear up their
connection further. We consider these two types of QKD are based
on the same essence, i.e., information splitting. To demonstrate
our viewpoint, we will introduce another QKD scheme, called the
mid protocol, and compare it with the above two types of QKD
respectively, which is similar to the way employed by Bennett
\emph{et al.} \cite{BBM92} when they analyzed the connection
between BB84 protocol and E91 protocol \cite{E91}.

To begin with, let us examine what results Alice obtains after the
information-splitting operation on the carrier in two-step-type
protocols. In Ref.\cite{GV95} (Ref.\cite{KI97}), with the help of
a Mach-Zehnder interferometer, the information unit is split into
two wave packets which have average (different) amplitudes. In
Refs.\cite{GLSLG,LL02,DLL05}, alternatively, the information
carrier is in a two-particle state and it is divided into two
single particles directly. It can be seen that, no matter how it
is split, both the two parts obtained are quantum ones in previous
schemes. Now we consider a special way to split, i.e., splitting
the information unit into one quantum part and one classical part.
A simple instance (i.e., the mid protocol) proceeds as follows
(also see Fig.~\ref{fig:one}).

(i) Alice generates a sequence of Einstein-Podolsky-Rosen (EPR)
pairs $(a_1,b_1)$, $(a_2,b_2)$, ..., $(a_m,b_m)$, where
$(a_i,b_i)$ denotes one EPR pair ($1\leq i\leq m$, $m>0$ is an
integer) and each pair is randomly in one of four Bell states,
i.e., $|\Phi^\pm\rangle=1/\sqrt{2}(|00\rangle\pm|11\rangle)$ and
$|\Psi^\pm\rangle=1/\sqrt{2}(|01\rangle\pm|10\rangle)$. Suppose
$|\Phi^+\rangle$, $|\Phi^-\rangle$, $|\Psi^+\rangle$, and
$|\Psi^-\rangle$ represent bit values 00, 01, 10, and 11,
respectively.

(ii) Alice splits each information unit into a quantum part and a
classical part. Here Alice takes every two adjoining pairs as a
complete carrier unit. Without loss of generality, consider the
first carrier unit $[a_1,b_1,a_2,b_2]$. The information-splitting
operation can be described as follows. Alice performs, at random,
one of the following two operations on the carrier unit: (1) doing
nothing; (2) exchanging the positions of particles $b_1$ and
$a_2$, that is, rearranging $[a_1,b_1,a_2,b_2]$ into
$[a_1,a_2,b_1,b_2]$, which can be easily realized by storage
rings. After that these particles form the quantum part. While the
classical part is 0 or 1, corresponding to the operation (1) or
(2) respectively.

(iii) Alice sends the two parts to Bob by two steps as in the
previous two-step-type protocols. More concretely, the quantum
part is sent firstly and then the classical part follows. The
delayed time is brought by the storage rings $SR_1$. That is, when
the quantum part is transmitted in the upper channel, the
classical part is still in $SR_1$. At the same time, the quantum
part is already in $SR_2$ when the classical part is transmitted
in the lower channel. As a result, Eve cannot access both parts
simultaneously.

(iv) When Bob have received both the two parts, he combines them
and tries to recover the complete information. If the classical
part is 1, Bob exchanges the positions of the second and the third
particles. Otherwise he does nothing.

(v) Bob measures this carrier unit to obtain the key. Bob performs
two Bell measurements on the first two particles and the rest two
particles. As a result, he can get four bits of key from his
measurement result (according to the above coding rule).

In the above description, for simplicity, we only considered the
first carrier unit, i.e., the first two EPR pairs. The other EPR
pairs generated by Alice can be used in a similar way. At last,
Alice and Bob publicly compare a subsequence of the bits sent by
Alice with those received by Bob to detect Eve's presence. If
there is no eavesdropping Alice and Bob will obtain a sequence of
secure key after error correction and privacy amplification
\cite{GRTZ}. The operation of order rearrangement is inspired by
the work of Deng and Long \cite{DL03}. It should be emphasized
that our main aim is using our scheme to discuss the relation
between BB84-type protocols and two-step-type ones instead of the
QKD scheme itself.

It can be seen that the mid protocol possesses the same features
of two-step-type protocols such as using orthogonal states as
carrier, splitting the information unit into two parts and sending
them one by one, etc. The only difference between the mid protocol
and two-step-type protocols is that the former splits the
information unit into a quantum part and a classical one, while
the latter splits it into two quantum parts. From this
perspective, the mid protocol is a variation of two-step-type
protocols.

Now let us observe the relation between the mid protocol and
BB84-type protocols. Obviously, if we interpret the delayed time
as the effect of storage rings, the delayed choice BB84-type
protocols can be depicted by Fig.~\ref{fig:one}, too.
Equivalently, Alice sends the photons to Bob through the upper
channel and, after Bob has received them, tells Bob the classical
bits (i.e., the bases information) through the lower channel.
Therefore, the mid protocol and BB84-type protocols have very
similar feature. Intuitively, the only difference between them is
that the mid protocol uses orthogonal states as carrier while
BB84-type protocols are based on nonorthogonal states. However, it
is just a superficial phenomenon. In fact, if we consider a
certain complete carrier unit such as $[a_1,b_1,a_2,b_2]$ in the
mid protocol, its possible states (i.e., the two possible states
after Alice's information-splitting operation) are nonorthogonal
indeed. For example, suppose both the two EPR pairs are in state
of $|\Phi^+\rangle$, that is,
$|\varphi\rangle_{a_1b_1}=|\varphi\rangle_{a_2b_2}=|\Phi^+\rangle$.
In this condition the state of this carrier unit can be written as
\begin{eqnarray}
|\varphi\rangle_{a_1b_1a_2b_2}&=&|\Phi^+\rangle\otimes|\Phi^+\rangle\\\nonumber
&=&\frac{1}{2}(|0000\rangle+|0011\rangle+|1100\rangle+|1111\rangle).
\end{eqnarray}
After Alice's information-splitting operation, this carrier unit
will be changed into one of two possible states
$|\varphi\rangle_{a_1b_1a_2b_2}$ and
$|\varphi\rangle_{a_1a_2b_1b_2}$, where
\begin{eqnarray}
|\varphi\rangle_{a_1a_2b_1b_2}=\frac{1}{2}(|0000\rangle+|0101\rangle+|1010\rangle+|1111\rangle).
\end{eqnarray}
From Eqs.(1) and (2) we can calculate the inner product of these
two states
\begin{eqnarray}
&&_{a_1b_1a_2b_2}\langle\varphi|\varphi\rangle_{a_1a_2b_1b_2}\\\nonumber
&=&\frac{1}{2}(\langle0000|+\langle0011|+\langle1100|+\langle1111|)\\\nonumber
&&\cdot\frac{1}{2}(|0000\rangle+|0101\rangle+|1010\rangle+|1111\rangle)=\frac{1}{2}.
\end{eqnarray}
Obviously, the two possible states are nonorthogonal and Eve can
never distinguish them determinately without Alice's classical
information. When the two EPR pairs are originally in other Bell
states, we can obtain the same conclusion by similar deduction.
Therefore, in substance, the states transmitted in the quantum
channel are nonorthogonal and the mid protocol has the same
essence as that of BB84-type protocols.

From another point of view, we can recognize the similarity
between the mid protocol and BB84-type protocols more clearly by
reinterpreting the latter. That is, Alice randomly prepares single
photons polarized along one of two possible directions, $0^\circ$
or $90^\circ$ (Note that they are orthogonal states). Afterwards,
Alice makes an information-splitting operation which can be
described as follows. Alice performs, at random, one of the
following two operations on each photon: (1) doing nothing; (2)
rotating its polarized direction by $45^\circ$ clockwise, that is,
changing its state into the corresponding one in the other basis
$\{45^\circ, 135^\circ\}$. After that the photon forms the quantum
part. While the classical part is 0 or 1, corresponding to the
operation (1) or (2) respectively. At the destination, when Bob
received both the two parts, he can recover and obtain the
original information just by a reverse operation and a measurement
in the basis $\{0^\circ, 90^\circ\}$. It can be seen that the
carrier states are orthogonal (nonorthogonal) before (after) the
information-splitting operation, which is quite similar with that
of the mid protocol. In a word, the mid protocol is actually
equivalent to BB84-type protocols (the difference between them is
their material carriers).

From above analysis we can draw a conclusion that BB84-type
protocols and two-step-type protocols are not entirely opposite.
Contrarily, they have the same essence, i.e., information
splitting. An orthogonal carrier or a nonorthogonal carrier, which
looks like the main difference between these two types of
protocols, is not an important, even not a very explicit matter.
Let us take the mid protocol as our example (BB84-type protocols
have the same feature). As far as Eve is concerned, the possible
states of the carrier unit, which is the smallest system she has
to distinguish, are nonorthogonal. But for Bob, the possible
states that need to be distinguished are indeed orthogonal because
Bob knows which two particles were initially in a Bell state (or
which basis was used when Alice prepared each photon in BB84-type
protocols) with the help of Alice's classical information. One may
argue that the possible states of the separate quantum parts in
some two-step-type protocols are not nonorthogonal. (For example,
the quantum part can be in mixed state, mostly in maximally mixed
state $\rho=\frac{1}{2}(|0\rangle\langle0|+|1\rangle\langle1|)$
\cite{GV95,LL02,DLL05}. Non-maximally mixed state is subtly
employed to remove the need for random timing tests in
Ref.\cite{KI97}.) However, we can find that each quantum part has
a feature that Eve cannot elicit the key information (i.e., the
complete information) without introducing disturbance, which is
same as that of the above so-called nonorthogonal states. Here we
generally call both of them ``indistinguishable states".

Before we conclude, it is worthwhile to inspect our so-called
``the same essence" of BB84-type protocols and two-step-type
protocols, which includes (1) the carrier state, which contains
the complete key information, is split into two parts (two quantum
parts or one quantum part and one classical one) and they are sent
one by one to prevent Eve from simultaneously taking control of
both of them. As described in Fig.~\ref{fig:one}, if Eve wants to
access both two parts at the same time, she must send fake
particles or wave packets to Bob, which will be detected by the
legal communicators; (2) the possible states of each quantum part
are indistinguishable states, which means that Eve cannot elicit
useful information without introducing errors. Of course, Eve
cannot extract key information from the classical part, either;
(3) anyone who obtains both parts can recover the initial carrier
state and then distinguish it from several possible (orthogonal or
distinguishable) states. Namely, when Bob received both two parts
he can obtain the key information Alice sent. In fact, the
security of BB84-type protocols and two-step-type protocols is
just based on the above essence.

Finally, we have to confess that there are still some differences
between BB84-type protocols and two-step-type protocols. The main
one is that the former splits the information unit into a quantum
part and a classical one, while the latter splits it into two
quantum parts. As a consequence, Bob can guess the value of the
classical part (i.e., the one comes later) instead of waiting
until its arriving in BB84-type protocols. Equivalently, Bob
randomly selects a basis to measure each photon and at last the
communicators discard the bits from those photons that are
prepared and measured in different bases, which is just the idea
of the original BB84 protocol \cite{BB84} (this feature also
applies to the mid protocol). On the contrary, in two-step-type
protocols the later part is a quantum one and Bob cannot do such
guesswork. It should be emphasized that, however, this difference
between these two types of QKD protocols cannot cover up their
same essence described above.

To summarize, we have mainly discussed the relation between
BB84-type protocols and two-step-type ones. By presenting a simple
QKD scheme (i.e., the mid protocol) we take cognizance of some
connections between these two types of protocols and draw a
conclusion that they have the same essence as described above (see
Fig.~\ref{fig:one}). This result can help us to make clear the
base of some QKD protocols' security. Furthermore, it is useful
for related scheme designing and security analyzing.

This work was supported by the National Natural Science Foundation
of China, Grant No. 60373059; the National Laboratory for Modern
Communications Science Foundation of China; the National Research
Foundation for the Doctoral Program of Higher Education of China,
Grant No.20040013007; the Graduate Students Innovation Foundation
of BUPT; and the ISN Open Foundation.

\end{document}